\documentclass[aps,prl,twocolumn,superscriptaddress]{revtex4-1}

\usepackage{graphicx}
\usepackage{amsmath}
\usepackage{bm}
\usepackage{xcolor}
\usepackage{multirow}
\usepackage{ulem}
\begin{document}


\title{Origin of negative electrocaloric effect in \textit{Pnma}-type antiferroelectric perovskites}


\author{Ningbo Fan}
\affiliation{Institute of Theoretical and Applied Physics, Soochow University, Suzhou 215006, China}
\affiliation{School of Physical Science and Technology, Soochow University, Suzhou 215006, China}

\author{Jorge \'{I}\~{n}iguez}
\affiliation{Materials Research and Technology Department, Luxembourg Institute of Science and Technology (LIST), Avenue des Hauts-Fourneaux 5, L-4362 Esch/Alzette, Luxembourg}
\affiliation{Department of Physics and Materials Science, University of Luxembourg, Rue du Brill 41, L-4422 Belvaux, Luxembourg}

\author{L. Bellaiche}
\affiliation{Physics Department and Institute for Nanoscience and Engineering, University of Arkansas, Fayetteville, Arkansas 72701, USA}

\author{Bin Xu}
\email[Email address: ]{binxu19@suda.edu.cn}
\affiliation{Institute of Theoretical and Applied Physics, Soochow University, Suzhou 215006, China}
\affiliation{School of Physical Science and Technology, Soochow University, Suzhou 215006, China}

\date{\today}

\begin{abstract}
Anomalous electrocaloric effect (ECE) with decreasing temperature upon application of an electric field is known to occur in antiferroelectrics (AFEs), and previous understanding refers to the field-induced canting of electric dipoles if there is no phase transitions. Here, we use a first-principle-based method to study the ECE in Nd-substituted BiFeO$_3$ (BNFO) perovskite solid solutions, which has the \textit{Pnma}-type AFE ground state. We demonstrate  another scenario to achieve and explain anomalous ECE, emphasizing that explicit consideration of octahedral tiltings is indispensable for a correct understanding. This mechanism may be general for AFEs for which the antipolar mode is not the primary order parameter. We also find that the negative ECE can reach a large magnitude in BNFO. 
\end{abstract}


\maketitle


Electrocaloric effect (ECE) can make temperature change via adiabatic application (or removal) of an electric field, providing an efficient approach for cooling or heating \cite{correlia2014,fahler2012,moya2014}. While ferroelectric (FE) or relaxor materials typically have ``normal'' positive sign of ECE, i.e., the temperature increases by applying a voltage, antiferroelectrics (AFEs) are known to have anomalous ECE that can yield an opposite sign \cite{pirc2014,geng2015,vales2021} . These two types of ECE can be utilized in combination to improve the performance of cooling/heating devices. 

Such negative (or inverse) caloric effect is also known to occur in other occasions, e.g.,  magnetic Heusler alloys, transitions between FE phases of different polarization directions, and application of an electric field against the polarization of a FE phase without switching \cite{gottschall2018,marathe2017,grunebohm2018}; however, its origin in AFEs is less well understood. AFEs materials are characterized by anti-polar atomic distortions that can be switched to a FE state under an electric field, and two mechanisms to explain their negative ECE with no AFE-FE transition have been proposed: 1) the ``dipole-canting'' model that dipolar entropy increases by misaligning the anti-parallel dipoles upon application of the field \cite{geng2015}; 2) the perturbative theory based on the Maxwell relation that only temperature and electric field dependencies of polarization need to be considered \cite{graf2021}. Interestingly, all these mechanisms only take the electric degrees of freedom explicitly into account. In contrast, most of the known AFEs are neither proper type (that is, the AFE phase is rarely driven by an AFE soft mode \cite{milesi2020}), nor systems with the anti-polar mode being the only significant order parameter. In fact, quite often, the AFE mode is secondary and coupled to other degrees of freedom, such as the octahedral tiltings in perovskites. For instance, PZO has a strong instability of anti-phase octahedral tilting \cite{xu2019}, while in \textit{Pnma}-type perovskite, such as rare earth orthoferrites and CsPbI$_3$, the anti-polar distortion arises from the condensation of both the in-plane anti-phase ($\omega_{R_{x,y}}$) and out-of-plane in-phase tiltings ($\omega_{M_z}$) via trilinear coupling \cite{bellaiche2013}. Although these tilting modes are non-polar, they couple strongly with the polar and anti-polar modes so that they can be influenced by the electric field as well, and in consequence contribute to the ECE.

To get a deeper understanding of the (negative) ECE in AFE, analysis based on Landau models involving the most relevant degrees of freedom have been proved to be very useful \cite{planes2014,zhang2011,ma2016,edstrom2020}, and it may thus be necessary that all the important order parameters are taken into consideration. Furthermore, some previous phenomenological models are often over simplified, since only one dimension is assumed \cite{pirc2014}. In reality, the direction of the applied field with respect to the crystallographic axis should have different effects regarding ECE. In this Letter, we take the antiferroelectric Nd substituted BiFeO$_3$ (BNFO) solid solution as an example and demonstrate that the octahedral tiltings can have very important effect on the sign and magnitude of the ECE. We also construct a phenomenological model that allows us to rationalize the contributions of each degree of freedom. In particular, the dipoles alone are found to  be insufficient to explain the negative ECE, while contributions from the in-phase and anti-phase tilting modes are indispensible. Moreover, BNFO is predicted to yield rather large negative ECE close to the AFE-to-FE transition.


BiFeO$_3$ (BFO) stabilizes in a \textit{R3c} ground state, but rare-earth doping with composition larger than 20$-$30\% is sufficient to alter it to the \textit{Pnma} structure \cite{karimi2009}. Here, we adopt the effective Hamiltonian scheme of Ref. \cite{xu2015} to study the Bi$_{0.6}$Nd$_{0.4}$FeO$_3$ solid solution under electric field at finite temperatures. With this composition, BNFO is stabilized in the AFE \textit{Pnma} phase at room temperature and can transform to a FE state under an electric  field \cite{karimi2009,karimi2009a,kan2010,xu2015,xu2017}. The solid solutions are simulated by a 12$\times$12$\times$12 supercell (containing 8,640 atoms) using Monte-Carlo (MC) simulations, in which the Bi and Nd atoms are randomly distributed (see Supplemental Material (SM) Section S1 for details \cite{sm}).


\begin{figure}[tbh]
	\includegraphics[width=0.80\linewidth]{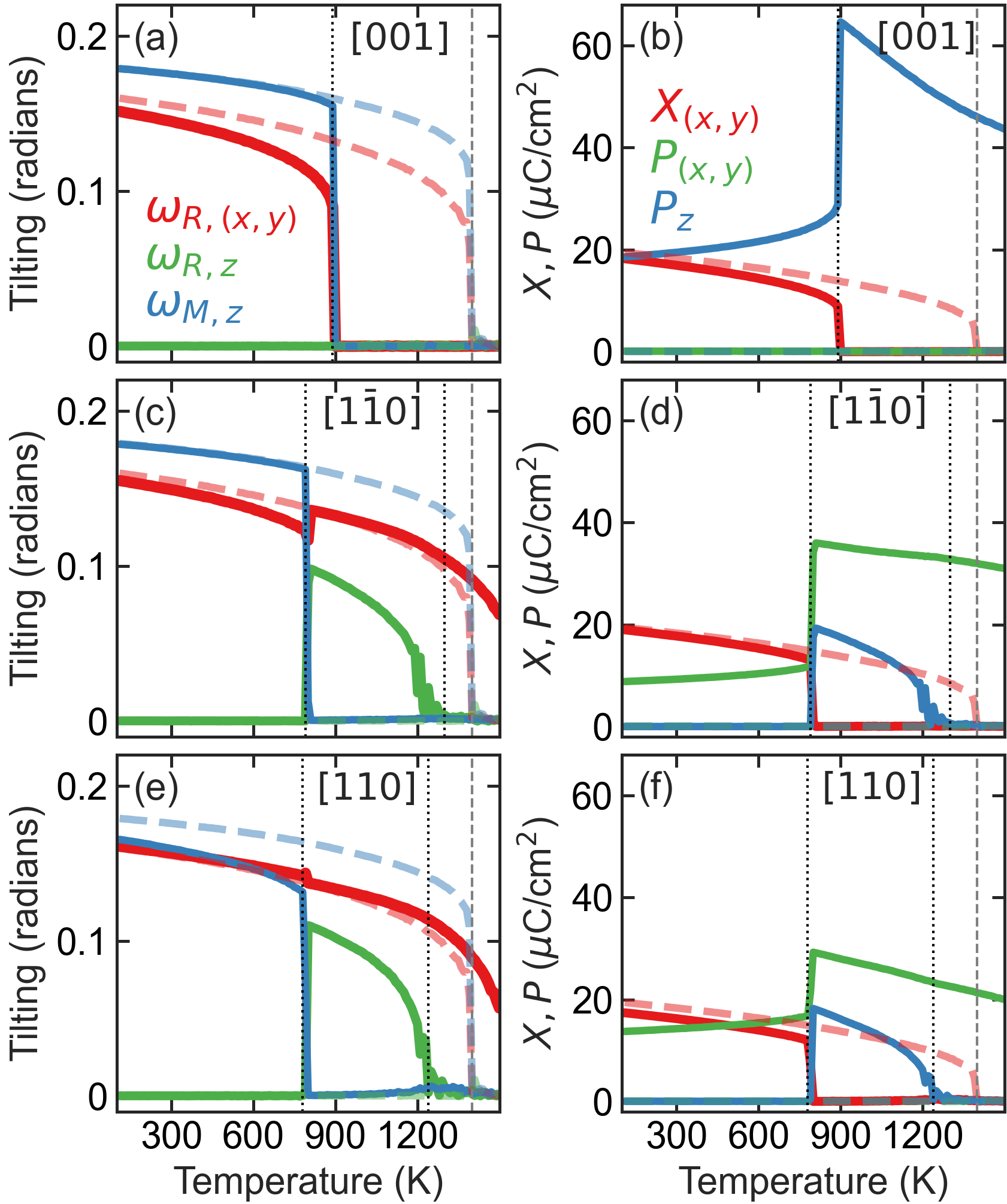}
	\caption{Effect of $\mathcal{E}$ field on the temperature dependence of the order parameters. (a),(b) $[001]$ field of 0.87 MV/cm (solid lines), in comparison with zero field (dashed lines), (c),(d) ibid for $[1\bar{1}0]$ field of 0.87 MV/cm, (e),(f) ibid for $[110]$ field of 0.61 MV/cm. The four relevant order parameters are anti-phase octahedral tilting $\omega_R$, in-phase tilting $\omega_M$, polarization $P$, and antiferroelectric vector $X$. Note that $P_x$=$-P_y$ for $[1\bar{1}0]$ field. The vertical dotted (or dashed) lines delimit different phases under finite (or zero) field.} \label{fig1}
\end{figure}

First, let us check how the order parameters are influenced by the electric field $\mathcal{E}$ \cite{1note}, using the effective Hamiltomian scheme. Note that the initial \textit{Pnma} structure has zero polarization, an in-plane anti-polar vector $X$ along the $[110]$ direction, and $a^-a^-c^+$ tilting in Glazer's notation (which corresponds to finite $x$- and $y$-components of the antiphase tilting vector, $\omega_{R,x}=\omega_{R,y}$, and finite $z$-component of the in-phase tilting vector, $\omega_{M,z}$). Three representative field directions are investigated (Fig. \ref{fig1}), together with the zero-field data (dashed curves) for comparison (for other $\mathcal{E}$ magnitudes, see SM Section S2 \cite{sm}). With no applied field, the \textit{Pnma} phase transforms to the paraelectric (PE) cubic phase at 1400 K (dashed lines in Fig. \ref{fig1}). 

Under $[001]$ field, we consider a representative case with $\mathcal{E}=$ 0.87 MV/cm, with which the AFE state transforms to the ferroelectric \textit{P4mm} phase at 880 K, characterized by a large polarization $P_z$ along $[001]$ and no octahedral tiltings. In the AFE state, one can see that the temperature dependence of $\omega_{M,z}$ does not differ much from the zero-field case, whereas $P$ and $\omega_{R,(x,y)}$ ($\omega_{R,x}=\omega_{R,y}$) show apparent changes. The moderate field-induced change of $X$ can be understood to a good approximation  via the change in $\omega_{R,(x,y)}$ since $X_{(x,y)}$ should be proportional to the product $\omega_{R,(x,y)}\omega_{M,z}$ -- as a result of a trilinear coupling between $X_{(x,y)}$, $\omega_{R,(x,y)}$ and $\omega_{M,z}$ \cite{xu2015hybrid}. Finite $P_z$ is induced by the field, whereas the field-induced suppression of $\omega_{R,(x,y)}$ is due to the competitive coupling with $P_z$. 

Moreover, if the field is applied along $[1\bar{1}0]$ (Figs. \ref{fig1}c and \ref{fig1}d), $P$ is induced along the same direction, i.e., a finite $P_x$=$-P_y$ first develops, and a transition to a FE \textit{Cc} phase occurs at 790 K, a structure characterized by a polarization in $\langle uuv\rangle$ direction ($u>v$) and $a^-a^-c^-$ octahedral tiltings ($a>c$). The third $\mathcal{E}$ direction is along $[110]$, along which $P$ develops, while $\omega_{M,z}$ is much suppressed and $\omega_{R,(x,y)}$ is more or less unchanged within the AFE-based state \cite{2note}. Such AFE phase then transforms also into a \textit{Cc} phase at 780 K (Fig. \ref{fig1}e and \ref{fig1}f).

Let us now concentrate on the ECE coefficient $\alpha=\frac{\partial T}{\partial \mathcal{E}}\rvert_S$ with $T$ being temperature and $S$ being entropy, which can be calculated from the cumulant formula using outputs of the MC simulations\cite{jiang2017electrocaloric,jiang2018giant,bin2016wang,jiang2021electrocaloric}.
\begin{align}\label{eq:alpha1}
	\alpha_{\text{MC}}=-Z^*a_{\text{latt}}T \left\{ \frac{\left< \left|\bm{u}\right|E_{\text{tot}} \right> - \left<\left|\bm{u}\right| \right> \left<E_{\text{tot}} \right>}{\left<E_{\text{tot}}^2 \right> - \left<E_{\text{tot}} \right> ^2 + \frac{21(k_{\text{B}}T)^2}{2N}}  \right\} \;,
\end{align}
where $Z^*$ is the Born effective charge associated with the local mode, $a_{\text{latt}}$ represents the lattice constant of the five-atom pseudo-cubic perovskite cell, $T$ is the simulation temperature, $\left|\bm{u}\right|$ is the supercell average of the magnitude of the local mode, $E_{\text{tot}}$ is the total energy given by the effective Hamiltonian, $k_\text{B}$ is the Boltzmann constant, $N$ is the number of sites in the supercell, and $\left< \; \right>$ denotes average over the MC sweeps at a given temperature.

\begin{figure}[tbh]
	\includegraphics[width=0.9\linewidth]{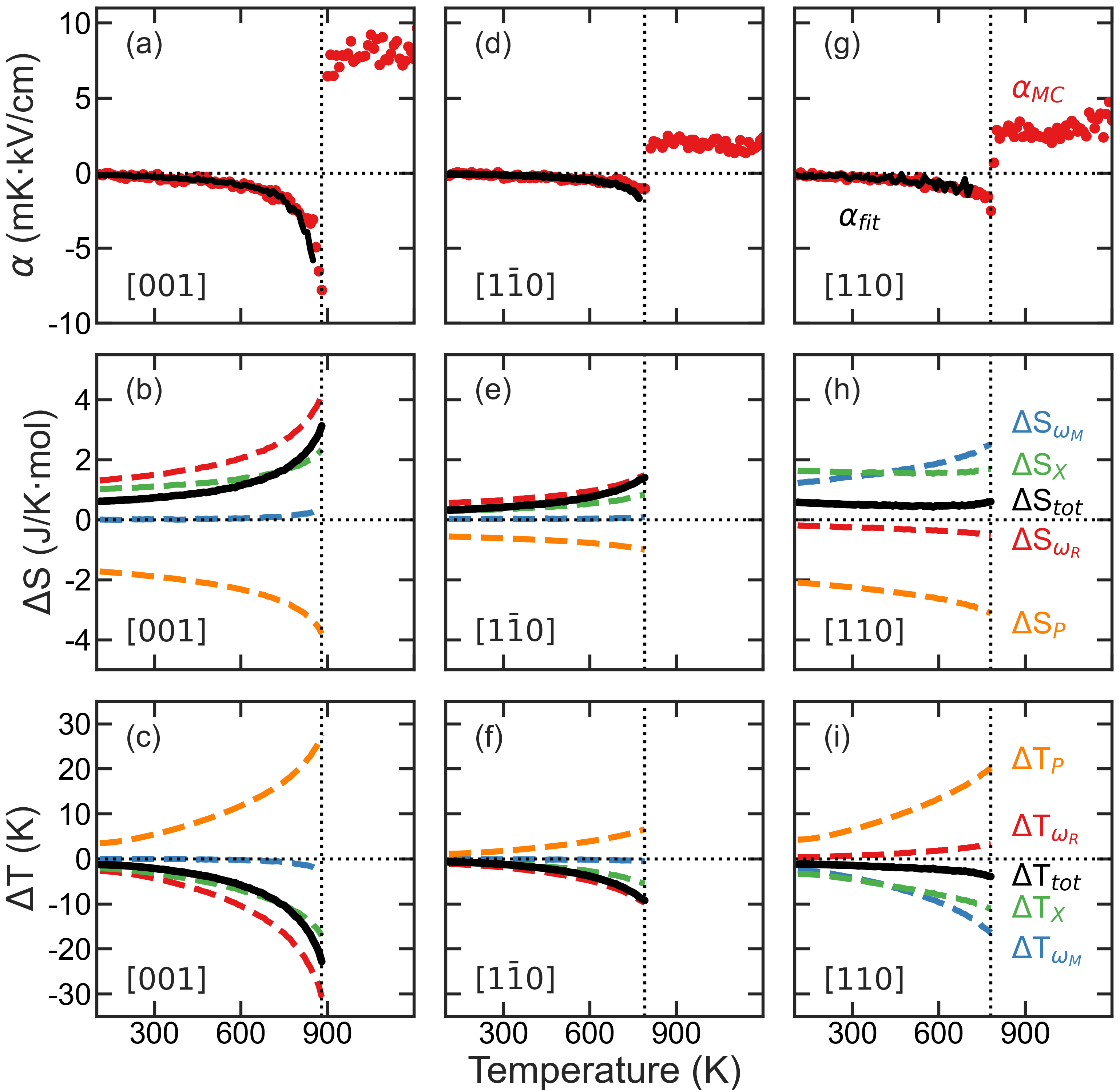}
	\caption{Temperature dependence of the calculated ECE coefficient $\alpha$, isothermal entropy change, and temperature change under (a)-(c) $[001]$ (0.87 MV/cm), (d)-(f) $[1\bar{1}0]$ (0.87 MV/cm), and (g),(i) $[110]$ (0.61 MV/cm) electric field. Colors in $\Delta S$ and $\Delta T$ denote the total and individual contributions from each order parameter. $\alpha_{\text{MC}}$ is computed by the cumulant approach, and $\alpha_{\text{fit}}$ is a fitted curve based on the Landau model.} \label{fig2}
\end{figure}

For fields applied along the $[001]$ direction (Fig. \ref{fig2}a), similar to the case in PZO-based AFE \cite{geng2015,vales2021}, $\alpha$ is negative in the AFE-based state, and its magnitude increases with temperature, which maximizes at the transition point where the AFE-based state disappears.  Across the phase transition, $\alpha$ jumps to be positive in the FE state, then (slightly) increases with temperature, as such qualitative temperature dependence is known for ferroelectrics \cite{jiang2021}. 

In order to have insightful analysis of the ECE,  Figs. \ref{fig2}b and \ref{fig2}c also report two other quantities related to electro-caloric response, namely the total isothermal change in entropy $\Delta S$, and the adiabatic temperature change $\Delta T$, as well as their individual contributions. 
Practically, in order to be able to compute the total entropy change, we consider the following Landau model for \textit{Pnma}-type AFE \cite{xu2015hybrid}.
\begin{align} \label{eq:landau}
	F=&F_0+\frac{1}{2}a_{\omega_R}(T)\omega_R^2 + \frac{1}{4}b_{\omega_R}\omega_R^4 + \frac{1}{2}a_{\omega_M}(T)\omega_M^2 +
	\frac{1}{4}b_{\omega_M}\omega_M^4  \nonumber \\
	&+ \frac{1}{2}a_X(T)X^2 + \frac{1}{4}b_XX^4 +
	\frac{1}{2}a_P(T)P^2 + \frac{1}{4}b_PP^4 \nonumber \\
	&  - \mathcal{E}P -
	cX\omega_R\omega_M + \frac{1}{2}d_1P^2\omega_R^2 + \frac{1}{2}d_2P^2\omega_M^2 \;.
\end{align}

Note that $F_0$ is the field-independent part of the free energy, only the quadratic coefficients $a_{\text{op}}=A_{\text{op}}(T-T_0^{\text{op}})$ have explicit temperature dependence, with ``op'' being the order parameter of $\omega_R$, $\omega_M$, $P$, or $X$, respectively, and $T_0^{\text{op}}$ is a transition temperature  for each order parameter. Also note that the tiltings are known to be the primary order parameters in many \textit{Pnma}-systems, while $X$ is secondary arising from the $ cX\omega_R\omega_M$ trilinear coupling involving in-phase and anti-phase tiltings \cite{zhao2014}. Recalling that $S=-\frac{\partial F}{\partial T}\rvert_{\mathcal{E},\text{op}}$, the change of entropy associated with a change of $\mathcal{E}$ can be written as:
\begin{align} \label{eq:entropy}
	\Delta S(T)=&-A_{\omega_R}\left[ \omega_R^2(T,\mathcal{E}_2) - \omega_R^2(T,\mathcal{E}_1) \right] \nonumber \\
	&-A_{\omega_M}\left[ \omega_M^2(T,\mathcal{E}_2) - \omega_M^2(T,\mathcal{E}_1) \right] \nonumber \\
	&-A_{P}\left[ P^2(T,\mathcal{E}_2) - P^2(T,\mathcal{E}_1) \right] \nonumber \\
	&-A_{X}\left[ X^2(T,\mathcal{E}_2) - X^2(T,\mathcal{E}_1) \right] \nonumber \\
	=&\Delta S_{\omega_R}+ \Delta S_{\omega_M} + \Delta S_P + \Delta S_X \; ,
\end{align}
where $\mathcal{E}_1$ and $\mathcal{E}_2$ are the initial and final electric field, respectively, and the values of the order parameters are obtained from MC simulations. The adiabatic temperature change due to ECE can be further obtained via $\Delta T=-(T/C_E)\Delta S$, where $C_E$ is the phonon specific heat \cite{kutnjak1999} -- which is calculated here by density functional theory (see SM Section S3 \cite{sm}). We thus have:
\begin{align} \label{eq:temperature}
	\Delta T=-\frac{T}{C_E}\Delta S 
	=\Delta T_{\omega_R}+ \Delta T_{\omega_M} + \Delta T_P + \Delta T_X \;.
\end{align}

Equations (\ref{eq:entropy}) and (\ref{eq:temperature}) are used to evaluate the entropy and temperature change (as well as their individual contributions associated with $\omega_R$, $\omega_M$, $P$ and $X$), in which each order parameter is directly obtained from MC simulations while the coefficients $A_{\text{op}}=-a_{\text{op}}^{\text{0-K}}/T_0^{\text{op}}$ are extracted by the following two steps: (1) finding $a_{\text{op}}^{\text{0-K}}$ from fitting the 0 K double-well energies based on our effective Hamiltonian to the Landau model (Eq. \ref{eq:landau}); (2) determining $T_0^{\text{op}}$ from fitting $\alpha$ calculated from the cumulant formula to that based on the Landau model with $\alpha_{\text{fit}}=\Delta T/(2\Delta \mathcal{E})$ for $\Delta \mathcal{E}$=0.02 MV/cm (see SM Section S4 for details\cite{sm}). Note that $\alpha_{\text{fit}}$ is shown in Fig. \ref{fig2}a and agrees rather well (over a large temperature range) with the $\alpha$ directly obtained from the MC simulations in the AFE-based state, therefore demonstrating the relevance of the aforementioned Landau model. Also note that the obtained $T_0^{\text{op}}$ reflects the depth of the energy wells.

One can then realize from Eq. (\ref{eq:temperature})  two important features for each $\Delta T_{\text{op}}$ associated with a specific order parameter (op), that are:  (1) such change is directly proportional to the $A_{\text{op}}$ coefficient related to that order parameter; and (2) such change involves the difference in the square of the order parameter between two electric fields, e.g., $\left[ \omega_R^2(T,\mathcal{E}_2) - \omega_R^2(T,\mathcal{E}_1) \right]$ for the anti-phase tilting. These four different individual changes in temperature are reported in Fig. \ref{fig2}c for field along $[001]$. 

Here, we only focus on the ECE in the AFE state, since latent heat needs to be considered when first-order AFE-to-FE transition occurs -- which is not included in Eq. \ref{eq:entropy}. Therefore, ECE involving the FE and PE states are not investigated within the scope of the present study. $\Delta S$ and $\Delta T$ in Figs. \ref{fig2}b and \ref{fig2}c, respectively, are practically calculated from Eqs. (\ref{eq:entropy}) and (\ref{eq:temperature})  with $\mathcal{E}_1$=0 and $\mathcal{E}_2$=0.87 MV/cm being along $[001]$. We numerically find that $\Delta T$ is negative and bears qualitatively similar $T$ dependence as that of $\alpha$, i.e., its magnitude increases with increasing temperature and maximizes at the transition point. On the other hand, $\Delta S$ has an opposite sign to that of $\Delta T$, as dictated by their relationship $\Delta T=-(T/C_E)\Delta S$. To understand how ECE is contributed by each order parameter, we can focus on the different $\Delta T_{\text{op}}$ displayed in Fig. \ref{fig2}c. For this $[001]$ field direction, $\omega_R$, $\omega_M$, and $X$ show negative contributions, with the magnitude from $\omega_R$ being the largest and that from $\omega_M$ being negligible, together with the rather large positive contribution from $P$. According to Eq. \ref{eq:temperature}, the negative (or positive) signs of the four different $\Delta T_{\text{op}}$ can be understood by the fact that the corresponding order parameter is suppressed (or enhanced) with the application of the field. In fact, as suggested by Eqs. (\ref{eq:entropy}) and (\ref{eq:temperature}), a large magnitude of $\Delta T_{\text{op}}$ relies on two factors, either those with large field-induced changes in their order parameter such as $\omega_R$ and $P$ here, or those with large coefficient $A_{\text{op}}$ such as $X$ here \cite{3note}, whereas $\Delta T_{\omega_M}$ for field along $[001]$ is negligible since both $A_{\omega_M}$ and the field-induced change in $\omega_M$ are very small.


\begin{figure*}[tbh]
	\includegraphics[width=0.9\linewidth]{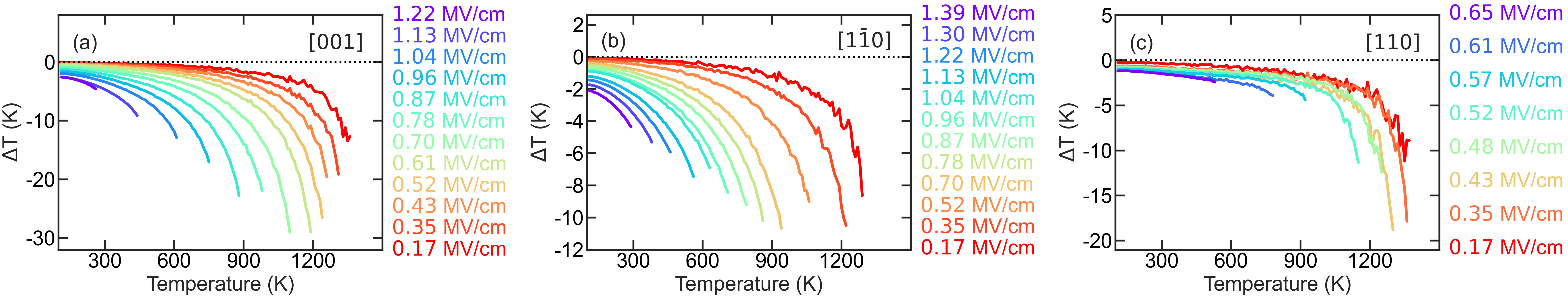}
	\caption{The ECE temperature change as a function of temperature under various electric field magnitude and directions. (a) $[001]$, (b) $[1\bar{1}0]$, and (c) $[110]$ field.} \label{fig3}
\end{figure*}

The same procedure is adopted to study the field directions along $[1\bar{1}0]$ and $[110]$. For the $[1\bar{1}0]$ field, as shown in Figs. \ref{fig2}d-f, a qualitatively similar ECE behavior is found compared with those from the $[001]$ field (namely, a negative $\alpha$ having a magnitude increasing when approaching the AFE-to-FE transition), except that the predicted $\alpha$, $\Delta S$ and $\Delta T$ are smaller in magnitude, e.g., the largest negative $\alpha$ = $-$1.0 mK$\cdot$kV/cm, $\Delta T$ = $-$9.2 K for the $[1\bar{1}0]$ field, to be compared with 
$-$7.8 mK$\cdot$kV/cm and $-$22.8 K for the $[001]$ field, respectively; however, the relative contributions of the order parameters share qualitatively similar characteristics, i.e., $\omega_R$ and $X$ are associated to negative $\Delta T_{\omega_R}$ and $\Delta T_{X}$, respectively, with $\omega_R$ contributing the most, together with the opposite effect from $P$ (positive $\Delta T_{P}$) and negligible contribution from $\omega_M$ (i.e., weak $\Delta T_{\omega_M}$). Note that $\alpha$ and $\Delta T$ are also negative and increase in magnitude with $T$ within the AFE-based state for the $[110]$ field (Fig. \ref{fig2}g-i), and their magnitude is about $-$2.5 mK$\cdot$kV/cm and $-$3.9 K, respectively, at the transition point.

However, albeit the qualitative similarities from the sign of $\Delta T_P$ and $\Delta T_X$ for all three $\mathcal{E}$ directions considered in Fig\sout{s}. \ref{fig2}, $\Delta T_{\omega_R}$ becomes positive  and $\Delta T_{\omega_M}$ shows large negative contribution only for $\mathcal{E}$ along $[110]$. Such unique features are associated to the small enhancement of $\omega_R$ and large suppression of $\omega_M$ under $[110]$ fields (see Fig. \ref{fig1}e). Further, the small ECE for $[110]$ field can be well understood by the fact that the negative $\Delta T_{\omega_M}$ and $\Delta T_{X}$ contributions are largely cancelled by the positive $\Delta T_{P}$ and $\Delta T_{\omega_R}$ contributions. It is also worth mentioning that $\Delta T_P$ is always positive and $\Delta T_X$ is always negative, as $P$ increases while $X$ decreases upon application of the $\mathcal{E}$ field (see Eq. \ref{eq:temperature}). In other words, being consistent with Eq. \ref{eq:entropy}, the applied field makes $P$ more ordered and $X$ more disordered.

It is interesting to note that the contrast of different ECE behaviors from various field directions are related to the relative orientation between $\mathcal{E}$ and $X$; for instance, $[001]$ and $[1\bar{1}0]$ fields are perpendicular to the antipolar vector, whereas the $[110]$ field is parallel to it. As detailed in SM Section S5 \cite{sm}, via the calculated energy landscape, we show numerically that minimization of the total energy dictates that increasing $P_z$ or $P_{(x,\bar{y})}$ in the \textit{Pnma} state would strongly suppress $\omega_R$, whereas such suppression is much less for $\omega_M$; and in contrast, the opposite is found for $P_{(x,y)}$, i.e., $\omega_M$ is strongly suppressed but $\omega_R$ is less correlated when $P_{(x,y)}$ increases. 

In fact, a continuous variation of $\Delta T_{\text{op}}$ can be realized if the $\mathcal{E}$ direction changes from $[110]$ towards $[001]$. In SM Section S6 \cite{sm}, we include results for $\mathcal{E}$ lying along $[111]$, $[112]$, $[113]$, and $[114]$ directions. While the overall $\Delta T$ remains negative and $\Delta T_{P}$ is always positive (because $P$ always increases with the field in the AFE-based phase), $\Delta T_{\omega_R}$ becomes more negative while $\Delta T_{\omega_M}$ becomes less negative,  as the angle between $\mathcal{E}$ and $X$  changes from 0 to 90 degrees. Nonetheless, it is worth to note that the contributions associate to $X$ and $P$ also show opposite trend with such rotation of the field, i.e., $\Delta T_P$ decreases (less positive) and $\Delta T_X$ increases (less negative). 

For each field direction, we can further change the field strength to tune the AFE transition temperature, where the maximum ECE occurs, thus manipulating the negative ECE. As depicted in Fig. \ref{fig3}, for $\mathcal{E}$ along $[001]$, $[1\bar{1}0]$ and $[110]$, the maximized negative $\Delta T$ is given by intermediate field, which allows the AFE-to-FE transition to occur at rather high temperature.  More precisely, if $\mathcal{E}$ is too small, the FE state is absent; if $\mathcal{E}$ is too large, the AFE-to-FE transition happens at low temperature and $\Delta T$ is relatively small. Remarkably, $\Delta T$ is predicted to reach $-$29.0 K  if $\mathcal{E}$ is along $[001]$ with a magnitude of 0.70 MV/cm at 1100 K, therefore being very large compared with the previously reported value of $-$5.8 K in La-doped Pb(Zr,Ti)O$_3$ \cite{geng2015}. However, if one is interested in a large negative ECE at {\it room temperature} (RT), a specific field direction and strength should be chosen, e.g., the largest negative $\Delta T$ of $-$4.7 K is realized at RT with a $[001]$ field having a magnitude of 1.13 MV/cm, with which the AFE-to-FE transition is close to RT. This value is comparable to that of PZO-based AFEs at RT \cite{geng2015,vales2021}.

Finally, let us examine the two previously proposed mechanisms. Regarding the ``dipole-canting'' mechanism \cite{geng2015}, in AFE BNFO, for all the investigated $\mathcal{E}$ directions, if only the dipole degrees of freedom (i.e., $P$ and $X$) are considered, canting of dipoles  indeed occurs in response to the field (Fig. \ref{fig1}). However, we numerically find that $\Delta T_P$ and $\Delta T_X$ have opposite signs, but the response of $P$ to $\mathcal{E}$ field is much greater than $X$ for all field directions. The corresponding change of temperature ($\Delta T_P+\Delta T_X$) is thus always positive, instead of negative, which therefore can not explain the overall negative ECE. This observation strongly suggests that explicit consideration of octahedral degrees of freedom is mandatory to correctly account for the (negative) ECE in AFEs. On the other hand, the perturbative approach solely based on polarization (as functions of $\mathcal{E}$ field and temperature) \cite{graf2021} agrees reasonably well with our predictions at low fields (see SM Section S7 \cite{sm}), indicating that the Maxwell relation $\frac{\partial S}{\partial \mathcal{E}}\rvert_T=\frac{\partial P}{\partial T}\rvert_\mathcal{E}$ holds for negative ECE systems with contributions from multiple degrees of freedom. 

In summary, our quantitative method show that, as octahedral tiltings are often primary order parameters in AFE perovskites, they have to be taken explicitly into account to construe the negative electrocaloric effect, and mere inclusion of dipolar degrees of freedom is insufficient. We also predict that AFE BNFO solid solutions are very promising to achieve giant temperature decrease near the AFE transition upon application of an electric field. Our atomistic mechanism also suggests that enhanced negative ECE may be possible upon freezing or suppression of order parameter(s) with positive $\Delta T_{\text{op}}$, for instance via interfacial proximity effect.

\begin{acknowledgments}
	This work is supported by the National Natural Science Foundation of China under Grant No. 12074277, Natural Science Foundation of Jiangsu Province (BK20201404), the startup fund from Soochow University and the support from Priority Academic Program Development (PAPD) of Jiangsu Higher Education Institutions. L.B. thanks the Office of Naval Research for the support under Grant No. N00014-21-1-2086.
\end{acknowledgments}



\begin{thebibliography}{53}%
\makeatletter
\providecommand \@ifxundefined [1]{%
 \@ifx{#1\undefined}
}%
\providecommand \@ifnum [1]{%
 \ifnum #1\expandafter \@firstoftwo
 \else \expandafter \@secondoftwo
 \fi
}%
\providecommand \@ifx [1]{%
 \ifx #1\expandafter \@firstoftwo
 \else \expandafter \@secondoftwo
 \fi
}%
\providecommand \natexlab [1]{#1}%
\providecommand \enquote  [1]{``#1''}%
\providecommand \bibnamefont  [1]{#1}%
\providecommand \bibfnamefont [1]{#1}%
\providecommand \citenamefont [1]{#1}%
\providecommand \href@noop [0]{\@secondoftwo}%
\providecommand \href [0]{\begingroup \@sanitize@url \@href}%
\providecommand \@href[1]{\@@startlink{#1}\@@href}%
\providecommand \@@href[1]{\endgroup#1\@@endlink}%
\providecommand \@sanitize@url [0]{\catcode `\\12\catcode `\$12\catcode
  `\&12\catcode `\#12\catcode `\^12\catcode `\_12\catcode `\%12\relax}%
\providecommand \@@startlink[1]{}%
\providecommand \@@endlink[0]{}%
\providecommand \url  [0]{\begingroup\@sanitize@url \@url }%
\providecommand \@url [1]{\endgroup\@href {#1}{\urlprefix }}%
\providecommand \urlprefix  [0]{URL }%
\providecommand \Eprint [0]{\href }%
\providecommand \doibase [0]{http://dx.doi.org/}%
\providecommand \selectlanguage [0]{\@gobble}%
\providecommand \bibinfo  [0]{\@secondoftwo}%
\providecommand \bibfield  [0]{\@secondoftwo}%
\providecommand \translation [1]{[#1]}%
\providecommand \BibitemOpen [0]{}%
\providecommand \bibitemStop [0]{}%
\providecommand \bibitemNoStop [0]{.\EOS\space}%
\providecommand \EOS [0]{\spacefactor3000\relax}%
\providecommand \BibitemShut  [1]{\csname bibitem#1\endcsname}%
\let\auto@bib@innerbib\@empty
\bibitem [{\citenamefont {Correlia}\ and\ \citenamefont
  {Zhang}(2014)}]{correlia2014}%
  \BibitemOpen
  \bibfield  {author} {\bibinfo {author} {\bibfnamefont {T.}~\bibnamefont
  {Correlia}}\ and\ \bibinfo {author} {\bibfnamefont {Q.}~\bibnamefont
  {Zhang}},\ }\href@noop {} {\emph {\bibinfo {title} {Electrocaloric materials:
  new generation of coolers}}}\ (\bibinfo  {publisher} {Springer, Berlin},\
  \bibinfo {year} {2014})\BibitemShut {NoStop}%
\bibitem [{\citenamefont {F{\"a}hler}\ \emph {et~al.}(2012)\citenamefont
  {F{\"a}hler}, \citenamefont {R{\"o}{\ss}ler}, \citenamefont {Kastner},
  \citenamefont {Eckert}, \citenamefont {Eggeler}, \citenamefont {Emmerich},
  \citenamefont {Entel}, \citenamefont {M{\"u}ller}, \citenamefont {Quandt},\
  and\ \citenamefont {Albe}}]{fahler2012}%
  \BibitemOpen
  \bibfield  {author} {\bibinfo {author} {\bibfnamefont {S.}~\bibnamefont
  {F{\"a}hler}}, \bibinfo {author} {\bibfnamefont {U.~K.}\ \bibnamefont
  {R{\"o}{\ss}ler}}, \bibinfo {author} {\bibfnamefont {O.}~\bibnamefont
  {Kastner}}, \bibinfo {author} {\bibfnamefont {J.}~\bibnamefont {Eckert}},
  \bibinfo {author} {\bibfnamefont {G.}~\bibnamefont {Eggeler}}, \bibinfo
  {author} {\bibfnamefont {H.}~\bibnamefont {Emmerich}}, \bibinfo {author}
  {\bibfnamefont {P.}~\bibnamefont {Entel}}, \bibinfo {author} {\bibfnamefont
  {S.}~\bibnamefont {M{\"u}ller}}, \bibinfo {author} {\bibfnamefont
  {E.}~\bibnamefont {Quandt}}, \ and\ \bibinfo {author} {\bibfnamefont
  {K.}~\bibnamefont {Albe}},\ }\href@noop {} {\bibfield  {journal} {\bibinfo
  {journal} {Adv. Eng. Mater.}\ }\textbf {\bibinfo {volume} {14}},\ \bibinfo
  {pages} {10} (\bibinfo {year} {2012})}\BibitemShut {NoStop}%
\bibitem [{\citenamefont {Moya}\ \emph {et~al.}(2014)\citenamefont {Moya},
  \citenamefont {Kar-Narayan},\ and\ \citenamefont {Mathur}}]{moya2014}%
  \BibitemOpen
  \bibfield  {author} {\bibinfo {author} {\bibfnamefont {X.}~\bibnamefont
  {Moya}}, \bibinfo {author} {\bibfnamefont {S.}~\bibnamefont {Kar-Narayan}}, \
  and\ \bibinfo {author} {\bibfnamefont {N.~D.}\ \bibnamefont {Mathur}},\
  }\href@noop {} {\bibfield  {journal} {\bibinfo  {journal} {Nat. Mater.}\
  }\textbf {\bibinfo {volume} {13}},\ \bibinfo {pages} {439} (\bibinfo {year}
  {2014})}\BibitemShut {NoStop}%
\bibitem [{\citenamefont {Pirc}\ \emph {et~al.}(2014)\citenamefont {Pirc},
  \citenamefont {Ro{\v{z}}i{\v{c}}}, \citenamefont {Koruza}, \citenamefont
  {Mali{\v{c}}},\ and\ \citenamefont {Kutnjak}}]{pirc2014}%
  \BibitemOpen
  \bibfield  {author} {\bibinfo {author} {\bibfnamefont {R.}~\bibnamefont
  {Pirc}}, \bibinfo {author} {\bibfnamefont {B.}~\bibnamefont
  {Ro{\v{z}}i{\v{c}}}}, \bibinfo {author} {\bibfnamefont {J.}~\bibnamefont
  {Koruza}}, \bibinfo {author} {\bibfnamefont {B.}~\bibnamefont {Mali{\v{c}}}},
  \ and\ \bibinfo {author} {\bibfnamefont {Z.}~\bibnamefont {Kutnjak}},\
  }\href@noop {} {\bibfield  {journal} {\bibinfo  {journal} {Europhys. Lett.}\
  }\textbf {\bibinfo {volume} {107}},\ \bibinfo {pages} {17002} (\bibinfo
  {year} {2014})}\BibitemShut {NoStop}%
\bibitem [{\citenamefont {Geng}\ \emph {et~al.}(2015)\citenamefont {Geng},
  \citenamefont {Liu}, \citenamefont {Meng}, \citenamefont {Bellaiche},
  \citenamefont {Scott}, \citenamefont {Dkhil},\ and\ \citenamefont
  {Jiang}}]{geng2015}%
  \BibitemOpen
  \bibfield  {author} {\bibinfo {author} {\bibfnamefont {W.}~\bibnamefont
  {Geng}}, \bibinfo {author} {\bibfnamefont {Y.}~\bibnamefont {Liu}}, \bibinfo
  {author} {\bibfnamefont {X.}~\bibnamefont {Meng}}, \bibinfo {author}
  {\bibfnamefont {L.}~\bibnamefont {Bellaiche}}, \bibinfo {author}
  {\bibfnamefont {J.~F.}\ \bibnamefont {Scott}}, \bibinfo {author}
  {\bibfnamefont {B.}~\bibnamefont {Dkhil}}, \ and\ \bibinfo {author}
  {\bibfnamefont {A.}~\bibnamefont {Jiang}},\ }\href@noop {} {\bibfield
  {journal} {\bibinfo  {journal} {Adv. Mater.}\ }\textbf {\bibinfo {volume}
  {27}},\ \bibinfo {pages} {3165} (\bibinfo {year} {2015})}\BibitemShut
  {NoStop}%
\bibitem [{\citenamefont {Vales-Castro}\ \emph {et~al.}(2021)\citenamefont
  {Vales-Castro}, \citenamefont {Faye}, \citenamefont {Vellvehi}, \citenamefont
  {Nouchokgwe}, \citenamefont {Perpi{\~n}{\`a}}, \citenamefont {Caicedo},
  \citenamefont {Jord{\`a}}, \citenamefont {Roleder}, \citenamefont {Kajewski},
  \citenamefont {Perez-Tomas} \emph {et~al.}}]{vales2021}%
  \BibitemOpen
  \bibfield  {author} {\bibinfo {author} {\bibfnamefont {P.}~\bibnamefont
  {Vales-Castro}}, \bibinfo {author} {\bibfnamefont {R.}~\bibnamefont {Faye}},
  \bibinfo {author} {\bibfnamefont {M.}~\bibnamefont {Vellvehi}}, \bibinfo
  {author} {\bibfnamefont {Y.}~\bibnamefont {Nouchokgwe}}, \bibinfo {author}
  {\bibfnamefont {X.}~\bibnamefont {Perpi{\~n}{\`a}}}, \bibinfo {author}
  {\bibfnamefont {J.~M.}\ \bibnamefont {Caicedo}}, \bibinfo {author}
  {\bibfnamefont {X.}~\bibnamefont {Jord{\`a}}}, \bibinfo {author}
  {\bibfnamefont {K.}~\bibnamefont {Roleder}}, \bibinfo {author} {\bibfnamefont
  {D.}~\bibnamefont {Kajewski}}, \bibinfo {author} {\bibfnamefont
  {A.}~\bibnamefont {Perez-Tomas}},  \emph {et~al.},\ }\href@noop {} {\bibfield
   {journal} {\bibinfo  {journal} {Phys. Rev. B}\ }\textbf {\bibinfo {volume}
  {103}},\ \bibinfo {pages} {054112} (\bibinfo {year} {2021})}\BibitemShut
  {NoStop}%
\bibitem [{\citenamefont {Gottschall}\ \emph {et~al.}(2018)\citenamefont
  {Gottschall}, \citenamefont {Gracia-Condal}, \citenamefont {Fries},
  \citenamefont {Taubel}, \citenamefont {Pfeuffer}, \citenamefont {Manosa},
  \citenamefont {Planes}, \citenamefont {Skokov},\ and\ \citenamefont
  {Gutfleisch}}]{gottschall2018}%
  \BibitemOpen
  \bibfield  {author} {\bibinfo {author} {\bibfnamefont {T.}~\bibnamefont
  {Gottschall}}, \bibinfo {author} {\bibfnamefont {A.}~\bibnamefont
  {Gracia-Condal}}, \bibinfo {author} {\bibfnamefont {M.}~\bibnamefont
  {Fries}}, \bibinfo {author} {\bibfnamefont {A.}~\bibnamefont {Taubel}},
  \bibinfo {author} {\bibfnamefont {L.}~\bibnamefont {Pfeuffer}}, \bibinfo
  {author} {\bibfnamefont {L.}~\bibnamefont {Manosa}}, \bibinfo {author}
  {\bibfnamefont {A.}~\bibnamefont {Planes}}, \bibinfo {author} {\bibfnamefont
  {K.~P.}\ \bibnamefont {Skokov}}, \ and\ \bibinfo {author} {\bibfnamefont
  {O.}~\bibnamefont {Gutfleisch}},\ }\href@noop {} {\bibfield  {journal}
  {\bibinfo  {journal} {Nat. Mater.}\ }\textbf {\bibinfo {volume} {17}},\
  \bibinfo {pages} {929} (\bibinfo {year} {2018})}\BibitemShut {NoStop}%
\bibitem [{\citenamefont {Marathe}\ \emph {et~al.}(2017)\citenamefont
  {Marathe}, \citenamefont {Renggli}, \citenamefont {Sanlialp}, \citenamefont
  {Karabasov}, \citenamefont {Shvartsman}, \citenamefont {Lupascu},
  \citenamefont {Gr{\"u}nebohm},\ and\ \citenamefont {Ederer}}]{marathe2017}%
  \BibitemOpen
  \bibfield  {author} {\bibinfo {author} {\bibfnamefont {M.}~\bibnamefont
  {Marathe}}, \bibinfo {author} {\bibfnamefont {D.}~\bibnamefont {Renggli}},
  \bibinfo {author} {\bibfnamefont {M.}~\bibnamefont {Sanlialp}}, \bibinfo
  {author} {\bibfnamefont {M.~O.}\ \bibnamefont {Karabasov}}, \bibinfo {author}
  {\bibfnamefont {V.~V.}\ \bibnamefont {Shvartsman}}, \bibinfo {author}
  {\bibfnamefont {D.~C.}\ \bibnamefont {Lupascu}}, \bibinfo {author}
  {\bibfnamefont {A.}~\bibnamefont {Gr{\"u}nebohm}}, \ and\ \bibinfo {author}
  {\bibfnamefont {C.}~\bibnamefont {Ederer}},\ }\href@noop {} {\bibfield
  {journal} {\bibinfo  {journal} {Phys. Rev. B}\ }\textbf {\bibinfo {volume}
  {96}},\ \bibinfo {pages} {014102} (\bibinfo {year} {2017})}\BibitemShut
  {NoStop}%
\bibitem [{\citenamefont {Gr{\"u}nebohm}\ \emph {et~al.}(2018)\citenamefont
  {Gr{\"u}nebohm}, \citenamefont {Ma}, \citenamefont {Marathe}, \citenamefont
  {Xu}, \citenamefont {Albe}, \citenamefont {Kalcher}, \citenamefont {Meyer},
  \citenamefont {Shvartsman}, \citenamefont {Lupascu},\ and\ \citenamefont
  {Ederer}}]{grunebohm2018}%
  \BibitemOpen
  \bibfield  {author} {\bibinfo {author} {\bibfnamefont {A.}~\bibnamefont
  {Gr{\"u}nebohm}}, \bibinfo {author} {\bibfnamefont {Y.-B.}\ \bibnamefont
  {Ma}}, \bibinfo {author} {\bibfnamefont {M.}~\bibnamefont {Marathe}},
  \bibinfo {author} {\bibfnamefont {B.-X.}\ \bibnamefont {Xu}}, \bibinfo
  {author} {\bibfnamefont {K.}~\bibnamefont {Albe}}, \bibinfo {author}
  {\bibfnamefont {C.}~\bibnamefont {Kalcher}}, \bibinfo {author} {\bibfnamefont
  {K.-C.}\ \bibnamefont {Meyer}}, \bibinfo {author} {\bibfnamefont {V.~V.}\
  \bibnamefont {Shvartsman}}, \bibinfo {author} {\bibfnamefont {D.~C.}\
  \bibnamefont {Lupascu}}, \ and\ \bibinfo {author} {\bibfnamefont
  {C.}~\bibnamefont {Ederer}},\ }\href@noop {} {\bibfield  {journal} {\bibinfo
  {journal} {Energy Technol.}\ }\textbf {\bibinfo {volume} {6}},\ \bibinfo
  {pages} {1491} (\bibinfo {year} {2018})}\BibitemShut {NoStop}%
\bibitem [{\citenamefont {Graf}\ and\ \citenamefont
  {{\'I}{\~n}iguez}(2021)}]{graf2021}%
  \BibitemOpen
  \bibfield  {author} {\bibinfo {author} {\bibfnamefont {M.}~\bibnamefont
  {Graf}}\ and\ \bibinfo {author} {\bibfnamefont {J.}~\bibnamefont
  {{\'I}{\~n}iguez}},\ }\href@noop {} {\bibfield  {journal} {\bibinfo
  {journal} {Commun. Mater}\ }\textbf {\bibinfo {volume} {2}},\ \bibinfo
  {pages} {1} (\bibinfo {year} {2021})}\BibitemShut {NoStop}%
\bibitem [{\citenamefont {Milesi-Brault}\ \emph {et~al.}(2020)\citenamefont
  {Milesi-Brault}, \citenamefont {Toulouse}, \citenamefont {Constable},
  \citenamefont {Aramberri}, \citenamefont {Simonet}, \citenamefont {De~Brion},
  \citenamefont {Berger}, \citenamefont {Paolasini}, \citenamefont {Bosak},
  \citenamefont {{\'I}{\~n}iguez} \emph {et~al.}}]{milesi2020}%
  \BibitemOpen
  \bibfield  {author} {\bibinfo {author} {\bibfnamefont {C.}~\bibnamefont
  {Milesi-Brault}}, \bibinfo {author} {\bibfnamefont {C.}~\bibnamefont
  {Toulouse}}, \bibinfo {author} {\bibfnamefont {E.}~\bibnamefont {Constable}},
  \bibinfo {author} {\bibfnamefont {H.}~\bibnamefont {Aramberri}}, \bibinfo
  {author} {\bibfnamefont {V.}~\bibnamefont {Simonet}}, \bibinfo {author}
  {\bibfnamefont {S.}~\bibnamefont {De~Brion}}, \bibinfo {author}
  {\bibfnamefont {H.}~\bibnamefont {Berger}}, \bibinfo {author} {\bibfnamefont
  {L.}~\bibnamefont {Paolasini}}, \bibinfo {author} {\bibfnamefont
  {A.}~\bibnamefont {Bosak}}, \bibinfo {author} {\bibfnamefont
  {J.}~\bibnamefont {{\'I}{\~n}iguez}},  \emph {et~al.},\ }\href@noop {}
  {\bibfield  {journal} {\bibinfo  {journal} {Phys. Rev. Lett.}\ }\textbf
  {\bibinfo {volume} {124}},\ \bibinfo {pages} {097603} (\bibinfo {year}
  {2020})}\BibitemShut {NoStop}%
\bibitem [{\citenamefont {Xu}\ \emph {et~al.}(2019)\citenamefont {Xu},
  \citenamefont {Hellman},\ and\ \citenamefont {Bellaiche}}]{xu2019}%
  \BibitemOpen
  \bibfield  {author} {\bibinfo {author} {\bibfnamefont {B.}~\bibnamefont
  {Xu}}, \bibinfo {author} {\bibfnamefont {O.}~\bibnamefont {Hellman}}, \ and\
  \bibinfo {author} {\bibfnamefont {L.}~\bibnamefont {Bellaiche}},\ }\href@noop
  {} {\bibfield  {journal} {\bibinfo  {journal} {Phys. Rev. B}\ }\textbf
  {\bibinfo {volume} {100}},\ \bibinfo {pages} {020102} (\bibinfo {year}
  {2019})}\BibitemShut {NoStop}%
\bibitem [{\citenamefont {Bellaiche}\ and\ \citenamefont
  {{\'I}{\~n}iguez}(2013)}]{bellaiche2013}%
  \BibitemOpen
  \bibfield  {author} {\bibinfo {author} {\bibfnamefont {L.}~\bibnamefont
  {Bellaiche}}\ and\ \bibinfo {author} {\bibfnamefont {J.}~\bibnamefont
  {{\'I}{\~n}iguez}},\ }\href@noop {} {\bibfield  {journal} {\bibinfo
  {journal} {Phys. Rev. B}\ }\textbf {\bibinfo {volume} {88}},\ \bibinfo
  {pages} {014104} (\bibinfo {year} {2013})}\BibitemShut {NoStop}%
\bibitem [{\citenamefont {Planes}\ \emph {et~al.}(2014)\citenamefont {Planes},
  \citenamefont {Castan},\ and\ \citenamefont {Saxena}}]{planes2014}%
  \BibitemOpen
  \bibfield  {author} {\bibinfo {author} {\bibfnamefont {A.}~\bibnamefont
  {Planes}}, \bibinfo {author} {\bibfnamefont {T.}~\bibnamefont {Castan}}, \
  and\ \bibinfo {author} {\bibfnamefont {A.}~\bibnamefont {Saxena}},\
  }\href@noop {} {\bibfield  {journal} {\bibinfo  {journal} {Philos. Mag.}\
  }\textbf {\bibinfo {volume} {94}},\ \bibinfo {pages} {1893} (\bibinfo {year}
  {2014})}\BibitemShut {NoStop}%
\bibitem [{\citenamefont {Zhang}\ \emph {et~al.}(2011)\citenamefont {Zhang},
  \citenamefont {Heitmann}, \citenamefont {Alpay},\ and\ \citenamefont
  {Rossetti~Jr}}]{zhang2011}%
  \BibitemOpen
  \bibfield  {author} {\bibinfo {author} {\bibfnamefont {J.}~\bibnamefont
  {Zhang}}, \bibinfo {author} {\bibfnamefont {A.}~\bibnamefont {Heitmann}},
  \bibinfo {author} {\bibfnamefont {S.}~\bibnamefont {Alpay}}, \ and\ \bibinfo
  {author} {\bibfnamefont {G.}~\bibnamefont {Rossetti~Jr}},\ }\href@noop {}
  {\bibfield  {journal} {\bibinfo  {journal} {Integr. Ferroelectr.}\ }\textbf
  {\bibinfo {volume} {125}},\ \bibinfo {pages} {168} (\bibinfo {year}
  {2011})}\BibitemShut {NoStop}%
\bibitem [{\citenamefont {Ma}\ \emph {et~al.}(2016)\citenamefont {Ma},
  \citenamefont {Novak}, \citenamefont {Koruza}, \citenamefont {Yang},
  \citenamefont {Albe},\ and\ \citenamefont {Xu}}]{ma2016}%
  \BibitemOpen
  \bibfield  {author} {\bibinfo {author} {\bibfnamefont {Y.-B.}\ \bibnamefont
  {Ma}}, \bibinfo {author} {\bibfnamefont {N.}~\bibnamefont {Novak}}, \bibinfo
  {author} {\bibfnamefont {J.}~\bibnamefont {Koruza}}, \bibinfo {author}
  {\bibfnamefont {T.}~\bibnamefont {Yang}}, \bibinfo {author} {\bibfnamefont
  {K.}~\bibnamefont {Albe}}, \ and\ \bibinfo {author} {\bibfnamefont {B.-X.}\
  \bibnamefont {Xu}},\ }\href@noop {} {\bibfield  {journal} {\bibinfo
  {journal} {Phys. Rev. B}\ }\textbf {\bibinfo {volume} {94}},\ \bibinfo
  {pages} {100104} (\bibinfo {year} {2016})}\BibitemShut {NoStop}%
\bibitem [{\citenamefont {Edstr{\"o}m}\ and\ \citenamefont
  {Ederer}(2020)}]{edstrom2020}%
  \BibitemOpen
  \bibfield  {author} {\bibinfo {author} {\bibfnamefont {A.}~\bibnamefont
  {Edstr{\"o}m}}\ and\ \bibinfo {author} {\bibfnamefont {C.}~\bibnamefont
  {Ederer}},\ }\href@noop {} {\bibfield  {journal} {\bibinfo  {journal} {Phys.
  Rev. Lett.}\ }\textbf {\bibinfo {volume} {124}},\ \bibinfo {pages} {167201}
  (\bibinfo {year} {2020})}\BibitemShut {NoStop}%
\bibitem [{\citenamefont {Karimi}\ \emph
  {et~al.}(2009{\natexlab{a}})\citenamefont {Karimi}, \citenamefont {Reaney},
  \citenamefont {Han}, \citenamefont {Pokorny},\ and\ \citenamefont
  {Sterianou}}]{karimi2009}%
  \BibitemOpen
  \bibfield  {author} {\bibinfo {author} {\bibfnamefont {S.}~\bibnamefont
  {Karimi}}, \bibinfo {author} {\bibfnamefont {I.}~\bibnamefont {Reaney}},
  \bibinfo {author} {\bibfnamefont {Y.}~\bibnamefont {Han}}, \bibinfo {author}
  {\bibfnamefont {J.}~\bibnamefont {Pokorny}}, \ and\ \bibinfo {author}
  {\bibfnamefont {I.}~\bibnamefont {Sterianou}},\ }\href@noop {} {\bibfield
  {journal} {\bibinfo  {journal} {J. Mater. Sci.}\ }\textbf {\bibinfo {volume}
  {44}},\ \bibinfo {pages} {5102} (\bibinfo {year}
  {2009}{\natexlab{a}})}\BibitemShut {NoStop}%
\bibitem [{\citenamefont {Xu}\ \emph {et~al.}(2015{\natexlab{a}})\citenamefont
  {Xu}, \citenamefont {Wang}, \citenamefont {\'I\~niguez},\ and\ \citenamefont
  {Bellaiche}}]{xu2015}%
  \BibitemOpen
  \bibfield  {author} {\bibinfo {author} {\bibfnamefont {B.}~\bibnamefont
  {Xu}}, \bibinfo {author} {\bibfnamefont {D.}~\bibnamefont {Wang}}, \bibinfo
  {author} {\bibfnamefont {J.}~\bibnamefont {\'I\~niguez}}, \ and\ \bibinfo
  {author} {\bibfnamefont {L.}~\bibnamefont {Bellaiche}},\ }\href {\doibase
  10.1002/adfm.201403811} {\bibfield  {journal} {\bibinfo  {journal} {Adv.
  Funct. Mater.}\ }\textbf {\bibinfo {volume} {25}},\ \bibinfo {pages} {552}
  (\bibinfo {year} {2015}{\natexlab{a}})}\BibitemShut {NoStop}%
\bibitem [{\citenamefont {Karimi}\ \emph
  {et~al.}(2009{\natexlab{b}})\citenamefont {Karimi}, \citenamefont {Reaney},
  \citenamefont {Levin},\ and\ \citenamefont {Sterianou}}]{karimi2009a}%
  \BibitemOpen
  \bibfield  {author} {\bibinfo {author} {\bibfnamefont {S.}~\bibnamefont
  {Karimi}}, \bibinfo {author} {\bibfnamefont {I.}~\bibnamefont {Reaney}},
  \bibinfo {author} {\bibfnamefont {I.}~\bibnamefont {Levin}}, \ and\ \bibinfo
  {author} {\bibfnamefont {I.}~\bibnamefont {Sterianou}},\ }\href@noop {}
  {\bibfield  {journal} {\bibinfo  {journal} {Appl. Phys. Lett.}\ }\textbf
  {\bibinfo {volume} {94}},\ \bibinfo {pages} {112903} (\bibinfo {year}
  {2009}{\natexlab{b}})}\BibitemShut {NoStop}%
\bibitem [{\citenamefont {Kan}\ \emph {et~al.}(2010)\citenamefont {Kan},
  \citenamefont {P{\'a}lov{\'a}}, \citenamefont {Anbusathaiah}, \citenamefont
  {Cheng}, \citenamefont {Fujino}, \citenamefont {Nagarajan}, \citenamefont
  {Rabe},\ and\ \citenamefont {Takeuchi}}]{kan2010}%
  \BibitemOpen
  \bibfield  {author} {\bibinfo {author} {\bibfnamefont {D.}~\bibnamefont
  {Kan}}, \bibinfo {author} {\bibfnamefont {L.}~\bibnamefont {P{\'a}lov{\'a}}},
  \bibinfo {author} {\bibfnamefont {V.}~\bibnamefont {Anbusathaiah}}, \bibinfo
  {author} {\bibfnamefont {C.~J.}\ \bibnamefont {Cheng}}, \bibinfo {author}
  {\bibfnamefont {S.}~\bibnamefont {Fujino}}, \bibinfo {author} {\bibfnamefont
  {V.}~\bibnamefont {Nagarajan}}, \bibinfo {author} {\bibfnamefont {K.~M.}\
  \bibnamefont {Rabe}}, \ and\ \bibinfo {author} {\bibfnamefont
  {I.}~\bibnamefont {Takeuchi}},\ }\href@noop {} {\bibfield  {journal}
  {\bibinfo  {journal} {Adv. Funct. Mater.}\ }\textbf {\bibinfo {volume}
  {20}},\ \bibinfo {pages} {1108} (\bibinfo {year} {2010})}\BibitemShut
  {NoStop}%
\bibitem [{\citenamefont {Xu}\ \emph {et~al.}(2017)\citenamefont {Xu},
  \citenamefont {{\'I}{\~n}iguez},\ and\ \citenamefont {Bellaiche}}]{xu2017}%
  \BibitemOpen
  \bibfield  {author} {\bibinfo {author} {\bibfnamefont {B.}~\bibnamefont
  {Xu}}, \bibinfo {author} {\bibfnamefont {J.}~\bibnamefont {{\'I}{\~n}iguez}},
  \ and\ \bibinfo {author} {\bibfnamefont {L.}~\bibnamefont {Bellaiche}},\
  }\href@noop {} {\bibfield  {journal} {\bibinfo  {journal} {Nat. Commun.}\
  }\textbf {\bibinfo {volume} {8}},\ \bibinfo {pages} {15682} (\bibinfo {year}
  {2017})}\BibitemShut {NoStop}%
\bibitem [{sm()}]{sm}%
  \BibitemOpen
  \href@noop {} {}\bibinfo {note} {See Supplemental Material for (1)
  Computational details, (2) Temperature dependencies of the order parameters
  for various fields, (3) DFT-calculated specific heat, (4) Determination of
  parameters in the Landau free-energy model, (5) Correlation between
  polarization and octahedral tiltings, (6) Additional $\mathcal{E}$-field
  directions from $[110]$ to $[001]$ and (7) Comparison with a perturbative
  approach, which additionally includes Refs.
  \cite{prosandeev2013,albrecht2010,kornev2007,lisenkov2009,zhong1994,zhong1995,kornev2006,daumont2012,kresse1996efficient,kresse1996efficiency,blochl1994projector,perdew1996generalized,baroni1987green,baroni2001phonons,gonze1989density,gonze1997dynamical,togo2015first,iniguez2014,wojdel2014}.}\BibitemShut
  {Stop}%
\bibitem [{1no()}]{1note}%
  \BibitemOpen
  \href@noop {} {}\bibinfo {note} {BNFO is multiferroic and in principle the
  magnetic moments can also respond to the applied electric field. Here it is
  not considered since we focus on the structural order parameters for a
  general AFE and numerically the magnetic contribution to $\alpha$ is found to
  the small (see Ref. \cite{jiang2021}).}\BibitemShut {Stop}%
\bibitem [{\citenamefont {Xu}\ \emph {et~al.}(2015{\natexlab{b}})\citenamefont
  {Xu}, \citenamefont {Wang}, \citenamefont {Zhao}, \citenamefont
  {{\'I}{\~n}iguez}, \citenamefont {Chen},\ and\ \citenamefont
  {Bellaiche}}]{xu2015hybrid}%
  \BibitemOpen
  \bibfield  {author} {\bibinfo {author} {\bibfnamefont {B.}~\bibnamefont
  {Xu}}, \bibinfo {author} {\bibfnamefont {D.}~\bibnamefont {Wang}}, \bibinfo
  {author} {\bibfnamefont {H.~J.}\ \bibnamefont {Zhao}}, \bibinfo {author}
  {\bibfnamefont {J.}~\bibnamefont {{\'I}{\~n}iguez}}, \bibinfo {author}
  {\bibfnamefont {X.~M.}\ \bibnamefont {Chen}}, \ and\ \bibinfo {author}
  {\bibfnamefont {L.}~\bibnamefont {Bellaiche}},\ }\href@noop {} {\bibfield
  {journal} {\bibinfo  {journal} {Adv. Funct. Mater.}\ }\textbf {\bibinfo
  {volume} {25}},\ \bibinfo {pages} {3626} (\bibinfo {year}
  {2015}{\natexlab{b}})}\BibitemShut {NoStop}%
\bibitem [{2no()}]{2note}%
  \BibitemOpen
  \href@noop {} {}\bibinfo {note} {Note that by AFE-based state, we mean the
  AFE state with all its associated order parameters but on top of which
  polarization develops along the field direction.}\BibitemShut {Stop}%
\bibitem [{\citenamefont {Jiang}\ \emph {et~al.}(2017)\citenamefont {Jiang},
  \citenamefont {Prokhorenko}, \citenamefont {Prosandeev}, \citenamefont
  {Nahas}, \citenamefont {Wang}, \citenamefont {{\'I}{\~n}iguez}, \citenamefont
  {Defay},\ and\ \citenamefont {Bellaiche}}]{jiang2017electrocaloric}%
  \BibitemOpen
  \bibfield  {author} {\bibinfo {author} {\bibfnamefont {Z.}~\bibnamefont
  {Jiang}}, \bibinfo {author} {\bibfnamefont {S.}~\bibnamefont {Prokhorenko}},
  \bibinfo {author} {\bibfnamefont {S.}~\bibnamefont {Prosandeev}}, \bibinfo
  {author} {\bibfnamefont {Y.}~\bibnamefont {Nahas}}, \bibinfo {author}
  {\bibfnamefont {D.}~\bibnamefont {Wang}}, \bibinfo {author} {\bibfnamefont
  {J.}~\bibnamefont {{\'I}{\~n}iguez}}, \bibinfo {author} {\bibfnamefont
  {E.}~\bibnamefont {Defay}}, \ and\ \bibinfo {author} {\bibfnamefont
  {L.}~\bibnamefont {Bellaiche}},\ }\href@noop {} {\bibfield  {journal}
  {\bibinfo  {journal} {Phys. Rev. B}\ }\textbf {\bibinfo {volume} {96}},\
  \bibinfo {pages} {014114} (\bibinfo {year} {2017})}\BibitemShut {NoStop}%
\bibitem [{\citenamefont {Jiang}\ \emph {et~al.}(2018)\citenamefont {Jiang},
  \citenamefont {Nahas}, \citenamefont {Prokhorenko}, \citenamefont
  {Prosandeev}, \citenamefont {Wang}, \citenamefont {{\'I}{\~n}iguez},\ and\
  \citenamefont {Bellaiche}}]{jiang2018giant}%
  \BibitemOpen
  \bibfield  {author} {\bibinfo {author} {\bibfnamefont {Z.}~\bibnamefont
  {Jiang}}, \bibinfo {author} {\bibfnamefont {Y.}~\bibnamefont {Nahas}},
  \bibinfo {author} {\bibfnamefont {S.}~\bibnamefont {Prokhorenko}}, \bibinfo
  {author} {\bibfnamefont {S.}~\bibnamefont {Prosandeev}}, \bibinfo {author}
  {\bibfnamefont {D.}~\bibnamefont {Wang}}, \bibinfo {author} {\bibfnamefont
  {J.}~\bibnamefont {{\'I}{\~n}iguez}}, \ and\ \bibinfo {author} {\bibfnamefont
  {L.}~\bibnamefont {Bellaiche}},\ }\href@noop {} {\bibfield  {journal}
  {\bibinfo  {journal} {Phys. Rev. B}\ }\textbf {\bibinfo {volume} {97}},\
  \bibinfo {pages} {104110} (\bibinfo {year} {2018})}\BibitemShut {NoStop}%
\bibitem [{\citenamefont {Bin-Omran}\ \emph {et~al.}(2016)\citenamefont
  {Bin-Omran}, \citenamefont {Kornev},\ and\ \citenamefont
  {Bellaiche}}]{bin2016wang}%
  \BibitemOpen
  \bibfield  {author} {\bibinfo {author} {\bibfnamefont {S.}~\bibnamefont
  {Bin-Omran}}, \bibinfo {author} {\bibfnamefont {I.~A.}\ \bibnamefont
  {Kornev}}, \ and\ \bibinfo {author} {\bibfnamefont {L.}~\bibnamefont
  {Bellaiche}},\ }\href@noop {} {\bibfield  {journal} {\bibinfo  {journal}
  {Phys. Rev. B}\ }\textbf {\bibinfo {volume} {93}},\ \bibinfo {pages} {014104}
  (\bibinfo {year} {2016})}\BibitemShut {NoStop}%
\bibitem [{\citenamefont {Jiang}\ \emph
  {et~al.}(2021{\natexlab{a}})\citenamefont {Jiang}, \citenamefont {Xu},
  \citenamefont {Prosandeev}, \citenamefont {Nahas}, \citenamefont
  {Prokhorenko}, \citenamefont {{\'I}{\~n}iguez},\ and\ \citenamefont
  {Bellaiche}}]{jiang2021electrocaloric}%
  \BibitemOpen
  \bibfield  {author} {\bibinfo {author} {\bibfnamefont {Z.}~\bibnamefont
  {Jiang}}, \bibinfo {author} {\bibfnamefont {B.}~\bibnamefont {Xu}}, \bibinfo
  {author} {\bibfnamefont {S.}~\bibnamefont {Prosandeev}}, \bibinfo {author}
  {\bibfnamefont {Y.}~\bibnamefont {Nahas}}, \bibinfo {author} {\bibfnamefont
  {S.}~\bibnamefont {Prokhorenko}}, \bibinfo {author} {\bibfnamefont
  {J.}~\bibnamefont {{\'I}{\~n}iguez}}, \ and\ \bibinfo {author} {\bibfnamefont
  {L.}~\bibnamefont {Bellaiche}},\ }\href@noop {} {\bibfield  {journal}
  {\bibinfo  {journal} {Phys. Rev. B}\ }\textbf {\bibinfo {volume} {103}},\
  \bibinfo {pages} {L100102} (\bibinfo {year}
  {2021}{\natexlab{a}})}\BibitemShut {NoStop}%
\bibitem [{\citenamefont {Jiang}\ \emph
  {et~al.}(2021{\natexlab{b}})\citenamefont {Jiang}, \citenamefont {Xu},
  \citenamefont {Prosandeev}, \citenamefont {Nahas}, \citenamefont
  {Prokhorenko}, \citenamefont {{\'I}{\~n}iguez},\ and\ \citenamefont
  {Bellaiche}}]{jiang2021}%
  \BibitemOpen
  \bibfield  {author} {\bibinfo {author} {\bibfnamefont {Z.}~\bibnamefont
  {Jiang}}, \bibinfo {author} {\bibfnamefont {B.}~\bibnamefont {Xu}}, \bibinfo
  {author} {\bibfnamefont {S.}~\bibnamefont {Prosandeev}}, \bibinfo {author}
  {\bibfnamefont {Y.}~\bibnamefont {Nahas}}, \bibinfo {author} {\bibfnamefont
  {S.}~\bibnamefont {Prokhorenko}}, \bibinfo {author} {\bibfnamefont
  {J.}~\bibnamefont {{\'I}{\~n}iguez}}, \ and\ \bibinfo {author} {\bibfnamefont
  {L.}~\bibnamefont {Bellaiche}},\ }\href@noop {} {\bibfield  {journal}
  {\bibinfo  {journal} {Phys. Rev. B}\ }\textbf {\bibinfo {volume} {103}},\
  \bibinfo {pages} {L100102} (\bibinfo {year}
  {2021}{\natexlab{b}})}\BibitemShut {NoStop}%
\bibitem [{\citenamefont {Zhao}\ \emph {et~al.}(2014)\citenamefont {Zhao},
  \citenamefont {{\'I}{\~n}iguez}, \citenamefont {Ren}, \citenamefont {Chen},\
  and\ \citenamefont {Bellaiche}}]{zhao2014}%
  \BibitemOpen
  \bibfield  {author} {\bibinfo {author} {\bibfnamefont {H.~J.}\ \bibnamefont
  {Zhao}}, \bibinfo {author} {\bibfnamefont {J.}~\bibnamefont
  {{\'I}{\~n}iguez}}, \bibinfo {author} {\bibfnamefont {W.}~\bibnamefont
  {Ren}}, \bibinfo {author} {\bibfnamefont {X.~M.}\ \bibnamefont {Chen}}, \
  and\ \bibinfo {author} {\bibfnamefont {L.}~\bibnamefont {Bellaiche}},\
  }\href@noop {} {\bibfield  {journal} {\bibinfo  {journal} {Phys. Rev. B}\
  }\textbf {\bibinfo {volume} {89}},\ \bibinfo {pages} {174101} (\bibinfo
  {year} {2014})}\BibitemShut {NoStop}%
\bibitem [{\citenamefont {Kutnjak}\ \emph {et~al.}(1999)\citenamefont
  {Kutnjak}, \citenamefont {Ro{\v{z}}i{\v{c}}},\ and\ \citenamefont
  {Pirc}}]{kutnjak1999}%
  \BibitemOpen
  \bibfield  {author} {\bibinfo {author} {\bibfnamefont {Z.}~\bibnamefont
  {Kutnjak}}, \bibinfo {author} {\bibfnamefont {B.}~\bibnamefont
  {Ro{\v{z}}i{\v{c}}}}, \ and\ \bibinfo {author} {\bibfnamefont
  {R.}~\bibnamefont {Pirc}},\ }\href@noop {} {\bibfield  {journal} {\bibinfo
  {journal} {Wiley encyclopedia of electrical and electronics engineering}\ ,\
  \bibinfo {pages} {1}} (\bibinfo {year} {1999})}\BibitemShut {NoStop}%
\bibitem [{3no()}]{3note}%
  \BibitemOpen
  \href@noop {} {}\bibinfo {note} {op with large coeﬀicient $A_{\text{op}}$
  is exptected to have strongly $T$-dependent change of the corresponding
  vibrational entropy.}\BibitemShut {Stop}%
\bibitem [{\citenamefont {Prosandeev}\ \emph {et~al.}(2013)\citenamefont
  {Prosandeev}, \citenamefont {Wang}, \citenamefont {Ren}, \citenamefont
  {{\'I}{\~n}iguez},\ and\ \citenamefont {Bellaiche}}]{prosandeev2013}%
  \BibitemOpen
  \bibfield  {author} {\bibinfo {author} {\bibfnamefont {S.}~\bibnamefont
  {Prosandeev}}, \bibinfo {author} {\bibfnamefont {D.}~\bibnamefont {Wang}},
  \bibinfo {author} {\bibfnamefont {W.}~\bibnamefont {Ren}}, \bibinfo {author}
  {\bibfnamefont {J.}~\bibnamefont {{\'I}{\~n}iguez}}, \ and\ \bibinfo {author}
  {\bibfnamefont {L.}~\bibnamefont {Bellaiche}},\ }\href@noop {} {\bibfield
  {journal} {\bibinfo  {journal} {Adv. Funct. Mater.}\ }\textbf {\bibinfo
  {volume} {23}},\ \bibinfo {pages} {234} (\bibinfo {year} {2013})}\BibitemShut
  {NoStop}%
\bibitem [{\citenamefont {Albrecht}\ \emph {et~al.}(2010)\citenamefont
  {Albrecht}, \citenamefont {Lisenkov}, \citenamefont {Ren}, \citenamefont
  {Rahmedov}, \citenamefont {Kornev},\ and\ \citenamefont
  {Bellaiche}}]{albrecht2010}%
  \BibitemOpen
  \bibfield  {author} {\bibinfo {author} {\bibfnamefont {D.}~\bibnamefont
  {Albrecht}}, \bibinfo {author} {\bibfnamefont {S.}~\bibnamefont {Lisenkov}},
  \bibinfo {author} {\bibfnamefont {W.}~\bibnamefont {Ren}}, \bibinfo {author}
  {\bibfnamefont {D.}~\bibnamefont {Rahmedov}}, \bibinfo {author}
  {\bibfnamefont {I.~A.}\ \bibnamefont {Kornev}}, \ and\ \bibinfo {author}
  {\bibfnamefont {L.}~\bibnamefont {Bellaiche}},\ }\href@noop {} {\bibfield
  {journal} {\bibinfo  {journal} {Phys. Rev. B}\ }\textbf {\bibinfo {volume}
  {81}},\ \bibinfo {pages} {140401} (\bibinfo {year} {2010})}\BibitemShut
  {NoStop}%
\bibitem [{\citenamefont {Kornev}\ \emph {et~al.}(2007)\citenamefont {Kornev},
  \citenamefont {Lisenkov}, \citenamefont {Haumont}, \citenamefont {Dkhil},\
  and\ \citenamefont {Bellaiche}}]{kornev2007}%
  \BibitemOpen
  \bibfield  {author} {\bibinfo {author} {\bibfnamefont {I.~A.}\ \bibnamefont
  {Kornev}}, \bibinfo {author} {\bibfnamefont {S.}~\bibnamefont {Lisenkov}},
  \bibinfo {author} {\bibfnamefont {R.}~\bibnamefont {Haumont}}, \bibinfo
  {author} {\bibfnamefont {B.}~\bibnamefont {Dkhil}}, \ and\ \bibinfo {author}
  {\bibfnamefont {L.}~\bibnamefont {Bellaiche}},\ }\href@noop {} {\bibfield
  {journal} {\bibinfo  {journal} {Phys. Rev. Lett.}\ }\textbf {\bibinfo
  {volume} {99}},\ \bibinfo {pages} {227602} (\bibinfo {year}
  {2007})}\BibitemShut {NoStop}%
\bibitem [{\citenamefont {Lisenkov}\ \emph {et~al.}(2009)\citenamefont
  {Lisenkov}, \citenamefont {Rahmedov},\ and\ \citenamefont
  {Bellaiche}}]{lisenkov2009}%
  \BibitemOpen
  \bibfield  {author} {\bibinfo {author} {\bibfnamefont {S.}~\bibnamefont
  {Lisenkov}}, \bibinfo {author} {\bibfnamefont {D.}~\bibnamefont {Rahmedov}},
  \ and\ \bibinfo {author} {\bibfnamefont {L.}~\bibnamefont {Bellaiche}},\
  }\href@noop {} {\bibfield  {journal} {\bibinfo  {journal} {Phys. Rev. Lett.}\
  }\textbf {\bibinfo {volume} {103}},\ \bibinfo {pages} {047204} (\bibinfo
  {year} {2009})}\BibitemShut {NoStop}%
\bibitem [{\citenamefont {Zhong}\ \emph {et~al.}(1994)\citenamefont {Zhong},
  \citenamefont {Vanderbilt},\ and\ \citenamefont {Rabe}}]{zhong1994}%
  \BibitemOpen
  \bibfield  {author} {\bibinfo {author} {\bibfnamefont {W.}~\bibnamefont
  {Zhong}}, \bibinfo {author} {\bibfnamefont {D.}~\bibnamefont {Vanderbilt}}, \
  and\ \bibinfo {author} {\bibfnamefont {K.~M.}\ \bibnamefont {Rabe}},\
  }\href@noop {} {\bibfield  {journal} {\bibinfo  {journal} {Phys. Rev. Lett.}\
  }\textbf {\bibinfo {volume} {73}},\ \bibinfo {pages} {1861} (\bibinfo {year}
  {1994})}\BibitemShut {NoStop}%
\bibitem [{\citenamefont {Zhong}\ and\ \citenamefont
  {Vanderbilt}(1995)}]{zhong1995}%
  \BibitemOpen
  \bibfield  {author} {\bibinfo {author} {\bibfnamefont {W.}~\bibnamefont
  {Zhong}}\ and\ \bibinfo {author} {\bibfnamefont {D.}~\bibnamefont
  {Vanderbilt}},\ }\href@noop {} {\bibfield  {journal} {\bibinfo  {journal}
  {Phys. Rev. Lett.}\ }\textbf {\bibinfo {volume} {74}},\ \bibinfo {pages}
  {2587} (\bibinfo {year} {1995})}\BibitemShut {NoStop}%
\bibitem [{\citenamefont {Kornev}\ \emph {et~al.}(2006)\citenamefont {Kornev},
  \citenamefont {Bellaiche}, \citenamefont {Janolin}, \citenamefont {Dkhil},\
  and\ \citenamefont {Suard}}]{kornev2006}%
  \BibitemOpen
  \bibfield  {author} {\bibinfo {author} {\bibfnamefont {I.~A.}\ \bibnamefont
  {Kornev}}, \bibinfo {author} {\bibfnamefont {L.}~\bibnamefont {Bellaiche}},
  \bibinfo {author} {\bibfnamefont {P.-E.}\ \bibnamefont {Janolin}}, \bibinfo
  {author} {\bibfnamefont {B.}~\bibnamefont {Dkhil}}, \ and\ \bibinfo {author}
  {\bibfnamefont {E.}~\bibnamefont {Suard}},\ }\href@noop {} {\bibfield
  {journal} {\bibinfo  {journal} {Phys. Rev. Lett.}\ }\textbf {\bibinfo
  {volume} {97}},\ \bibinfo {pages} {157601} (\bibinfo {year}
  {2006})}\BibitemShut {NoStop}%
\bibitem [{\citenamefont {Daumont}\ \emph {et~al.}(2012)\citenamefont
  {Daumont}, \citenamefont {Ren}, \citenamefont {Infante}, \citenamefont
  {Lisenkov}, \citenamefont {Allibe}, \citenamefont {Carr{\'e}t{\'e}ro},
  \citenamefont {Fusil}, \citenamefont {Jacquet}, \citenamefont {Bouvet},
  \citenamefont {Bouamrane} \emph {et~al.}}]{daumont2012}%
  \BibitemOpen
  \bibfield  {author} {\bibinfo {author} {\bibfnamefont {C.}~\bibnamefont
  {Daumont}}, \bibinfo {author} {\bibfnamefont {W.}~\bibnamefont {Ren}},
  \bibinfo {author} {\bibfnamefont {I.}~\bibnamefont {Infante}}, \bibinfo
  {author} {\bibfnamefont {S.}~\bibnamefont {Lisenkov}}, \bibinfo {author}
  {\bibfnamefont {J.}~\bibnamefont {Allibe}}, \bibinfo {author} {\bibfnamefont
  {C.}~\bibnamefont {Carr{\'e}t{\'e}ro}}, \bibinfo {author} {\bibfnamefont
  {S.}~\bibnamefont {Fusil}}, \bibinfo {author} {\bibfnamefont
  {E.}~\bibnamefont {Jacquet}}, \bibinfo {author} {\bibfnamefont
  {T.}~\bibnamefont {Bouvet}}, \bibinfo {author} {\bibfnamefont
  {F.}~\bibnamefont {Bouamrane}},  \emph {et~al.},\ }\href@noop {} {\bibfield
  {journal} {\bibinfo  {journal} {J. Phys.: Condens. Matter}\ }\textbf
  {\bibinfo {volume} {24}},\ \bibinfo {pages} {162202} (\bibinfo {year}
  {2012})}\BibitemShut {NoStop}%
\bibitem [{\citenamefont {Kresse}\ and\ \citenamefont
  {Furthm{\"u}ller}(1996{\natexlab{a}})}]{kresse1996efficient}%
  \BibitemOpen
  \bibfield  {author} {\bibinfo {author} {\bibfnamefont {G.}~\bibnamefont
  {Kresse}}\ and\ \bibinfo {author} {\bibfnamefont {J.}~\bibnamefont
  {Furthm{\"u}ller}},\ }\href@noop {} {\bibfield  {journal} {\bibinfo
  {journal} {Phys. Rev. B}\ }\textbf {\bibinfo {volume} {54}},\ \bibinfo
  {pages} {11169} (\bibinfo {year} {1996}{\natexlab{a}})}\BibitemShut {NoStop}%
\bibitem [{\citenamefont {Kresse}\ and\ \citenamefont
  {Furthm{\"u}ller}(1996{\natexlab{b}})}]{kresse1996efficiency}%
  \BibitemOpen
  \bibfield  {author} {\bibinfo {author} {\bibfnamefont {G.}~\bibnamefont
  {Kresse}}\ and\ \bibinfo {author} {\bibfnamefont {J.}~\bibnamefont
  {Furthm{\"u}ller}},\ }\href@noop {} {\bibfield  {journal} {\bibinfo
  {journal} {Comput. Mater. Sci}\ }\textbf {\bibinfo {volume} {6}},\ \bibinfo
  {pages} {15} (\bibinfo {year} {1996}{\natexlab{b}})}\BibitemShut {NoStop}%
\bibitem [{\citenamefont {Bl{\"o}chl}(1994)}]{blochl1994projector}%
  \BibitemOpen
  \bibfield  {author} {\bibinfo {author} {\bibfnamefont {P.~E.}\ \bibnamefont
  {Bl{\"o}chl}},\ }\href@noop {} {\bibfield  {journal} {\bibinfo  {journal}
  {Phys. Rev. B}\ }\textbf {\bibinfo {volume} {50}},\ \bibinfo {pages} {17953}
  (\bibinfo {year} {1994})}\BibitemShut {NoStop}%
\bibitem [{\citenamefont {Perdew}\ \emph {et~al.}(1996)\citenamefont {Perdew},
  \citenamefont {Burke},\ and\ \citenamefont
  {Ernzerhof}}]{perdew1996generalized}%
  \BibitemOpen
  \bibfield  {author} {\bibinfo {author} {\bibfnamefont {J.~P.}\ \bibnamefont
  {Perdew}}, \bibinfo {author} {\bibfnamefont {K.}~\bibnamefont {Burke}}, \
  and\ \bibinfo {author} {\bibfnamefont {M.}~\bibnamefont {Ernzerhof}},\
  }\href@noop {} {\bibfield  {journal} {\bibinfo  {journal} {Phys. Rev. Lett.}\
  }\textbf {\bibinfo {volume} {77}},\ \bibinfo {pages} {3865} (\bibinfo {year}
  {1996})}\BibitemShut {NoStop}%
\bibitem [{\citenamefont {Baroni}\ \emph {et~al.}(1987)\citenamefont {Baroni},
  \citenamefont {Giannozzi},\ and\ \citenamefont {Testa}}]{baroni1987green}%
  \BibitemOpen
  \bibfield  {author} {\bibinfo {author} {\bibfnamefont {S.}~\bibnamefont
  {Baroni}}, \bibinfo {author} {\bibfnamefont {P.}~\bibnamefont {Giannozzi}}, \
  and\ \bibinfo {author} {\bibfnamefont {A.}~\bibnamefont {Testa}},\
  }\href@noop {} {\bibfield  {journal} {\bibinfo  {journal} {Phys. Rev. Lett.}\
  }\textbf {\bibinfo {volume} {58}},\ \bibinfo {pages} {1861} (\bibinfo {year}
  {1987})}\BibitemShut {NoStop}%
\bibitem [{\citenamefont {Baroni}\ \emph {et~al.}(2001)\citenamefont {Baroni},
  \citenamefont {De~Gironcoli}, \citenamefont {Dal~Corso},\ and\ \citenamefont
  {Giannozzi}}]{baroni2001phonons}%
  \BibitemOpen
  \bibfield  {author} {\bibinfo {author} {\bibfnamefont {S.}~\bibnamefont
  {Baroni}}, \bibinfo {author} {\bibfnamefont {S.}~\bibnamefont
  {De~Gironcoli}}, \bibinfo {author} {\bibfnamefont {A.}~\bibnamefont
  {Dal~Corso}}, \ and\ \bibinfo {author} {\bibfnamefont {P.}~\bibnamefont
  {Giannozzi}},\ }\href@noop {} {\bibfield  {journal} {\bibinfo  {journal}
  {Rev. Mod. Phys.}\ }\textbf {\bibinfo {volume} {73}},\ \bibinfo {pages} {515}
  (\bibinfo {year} {2001})}\BibitemShut {NoStop}%
\bibitem [{\citenamefont {Gonze}\ and\ \citenamefont
  {Vigneron}(1989)}]{gonze1989density}%
  \BibitemOpen
  \bibfield  {author} {\bibinfo {author} {\bibfnamefont {X.}~\bibnamefont
  {Gonze}}\ and\ \bibinfo {author} {\bibfnamefont {J.-P.}\ \bibnamefont
  {Vigneron}},\ }\href@noop {} {\bibfield  {journal} {\bibinfo  {journal}
  {Phys. Rev. B}\ }\textbf {\bibinfo {volume} {39}},\ \bibinfo {pages} {13120}
  (\bibinfo {year} {1989})}\BibitemShut {NoStop}%
\bibitem [{\citenamefont {Gonze}\ and\ \citenamefont
  {Lee}(1997)}]{gonze1997dynamical}%
  \BibitemOpen
  \bibfield  {author} {\bibinfo {author} {\bibfnamefont {X.}~\bibnamefont
  {Gonze}}\ and\ \bibinfo {author} {\bibfnamefont {C.}~\bibnamefont {Lee}},\
  }\href@noop {} {\bibfield  {journal} {\bibinfo  {journal} {Phys. Rev. B}\
  }\textbf {\bibinfo {volume} {55}},\ \bibinfo {pages} {10355} (\bibinfo {year}
  {1997})}\BibitemShut {NoStop}%
\bibitem [{\citenamefont {Togo}\ and\ \citenamefont
  {Tanaka}(2015)}]{togo2015first}%
  \BibitemOpen
  \bibfield  {author} {\bibinfo {author} {\bibfnamefont {A.}~\bibnamefont
  {Togo}}\ and\ \bibinfo {author} {\bibfnamefont {I.}~\bibnamefont {Tanaka}},\
  }\href@noop {} {\bibfield  {journal} {\bibinfo  {journal} {Scr. Mater.}\
  }\textbf {\bibinfo {volume} {108}},\ \bibinfo {pages} {1} (\bibinfo {year}
  {2015})}\BibitemShut {NoStop}%
\bibitem [{\citenamefont {{\'I}{\~n}iguez}\ \emph {et~al.}(2014)\citenamefont
  {{\'I}{\~n}iguez}, \citenamefont {Stengel}, \citenamefont {Prosandeev},\ and\
  \citenamefont {Bellaiche}}]{iniguez2014}%
  \BibitemOpen
  \bibfield  {author} {\bibinfo {author} {\bibfnamefont {J.}~\bibnamefont
  {{\'I}{\~n}iguez}}, \bibinfo {author} {\bibfnamefont {M.}~\bibnamefont
  {Stengel}}, \bibinfo {author} {\bibfnamefont {S.}~\bibnamefont {Prosandeev}},
  \ and\ \bibinfo {author} {\bibfnamefont {L.}~\bibnamefont {Bellaiche}},\
  }\href@noop {} {\bibfield  {journal} {\bibinfo  {journal} {Phys. Rev. B}\
  }\textbf {\bibinfo {volume} {90}},\ \bibinfo {pages} {220103} (\bibinfo
  {year} {2014})}\BibitemShut {NoStop}%
\bibitem [{\citenamefont {Wojdeł}\ and\ \citenamefont
  {{\'I}{\~n}iguez}(2014)}]{wojdel2014}%
  \BibitemOpen
  \bibfield  {author} {\bibinfo {author} {\bibfnamefont {J.~C.}\ \bibnamefont
  {Wojde{\l}}}\ and\ \bibinfo {author} {\bibfnamefont {J.}~\bibnamefont
  {{\'I}{\~n}iguez}},\ }\href@noop {} {\bibfield  {journal} {\bibinfo
  {journal} {Phys. Rev. B}\ }\textbf {\bibinfo {volume} {90}},\ \bibinfo
  {pages} {014105} (\bibinfo {year} {2014})}\BibitemShut {NoStop}%
\end{thebibliography}

%

\end{document}